\documentclass[12pt,a4paper]{article}

\usepackage{graphicx}

\usepackage{ifpdf}  
\ifpdf
\usepackage[pdftex,bookmarks=true,bookmarksnumbered=false,breaklinks=true]{hyperref}
\DeclareGraphicsExtensions{.pdf,.png,.jpg}

\else

\usepackage[ps2pdf,linktocpage=true,bookmarks=true,bookmarksnumbered=false,breaklinks=true]{hyperref}
\DeclareGraphicsExtensions{.eps}

\fi

\usepackage{booktabs}

\usepackage{amsmath,amsfonts}
\usepackage{amsthm,amssymb}
\usepackage{caption} % although gives warning of unsopported package.
\usepackage{subcaption} % although gives warning of unsopported package.
\providecommand{\keywords}[1]
{
	\small	
	\textbf{\textit{Keywords---}} #1
}

\theoremstyle{remark}

\theoremstyle{definition}  
\newtheorem{definition}{Definition} 
 
\newtheorem{assumption}{Assumption} 

\newcommand{\R}{\mathbb {R}}
\newcommand{\FF}{F_{t^{+}}}
\newcommand{\FFF}{F_{{t-1}^{+}}}
\newcommand{\FFFF}{F_{{t+1}^{+}}}
\newcommand{\sS}{\mathcal {S}}
\newcommand{\E}{\mathbb {E}}

\newcommand{\tabref}[1]{Table~\ref{#1}}
\newcommand{\figref}[1]{Figure~\ref{#1}}

\newcommand{\abs}[1]{\left\vert {#1} \right\vert}

\begin{document}
	
	\title{The QLBS Model within the presence of feedback loops through the impacts of a large trader}
	 \author{Ahmet Umur Özsoy$^{1,2}$, Ömür Uğur$^{2}$  \\
	 	{\footnotesize ${}^{1}$Turkish Management Sciences Institute (TÜSSİDE), TÜBİTAK}\\ 
	 	{\footnotesize 41401 Kocaeli, Turkey (E-Mail: \texttt{umurozsoy@gmail.com})} \\
	 	{\footnotesize ${}^{2}$Middle East Technical University, Institute of Applied Mathematics}\\
	 	{\footnotesize 06800 Ankara, Turkey (E-Mail: \texttt{ougur@metu.edu.tr})}
	 }
	\date{}
	
	\maketitle
	
\begin{abstract} % is it too wordy? 
	We extend the QLBS model by reformulating via considering a large trader 
	whose transactions leave a permanent impact on the evolution 
	of the exchange rate process and therefore affect the price 
	of contingent claims on such processes. 
	Through a hypothetical limit order book we quantify the exchange rate 
	altered by such transactions. 
	We therefore define the quoted exchange rate process, 
	for which we assume the existence of a postulated hedging strategy. 
	Given the quoted exchange rate and postulated hedging strategy, 
	we find an optimal hedging strategy through 
	batch-mode reinforcement learning given the trader alters 
	the course of the exchange rate process. 
	We assume that the trader has its own concept of fair price 
	and we define our problem as finding the hedging strategy 
	with much lower transaction costs yet delivering a price 
	that well converges to the fair price of the trader. 
	We show our contribution results in an optimal hedging strategy 
	with much lower transaction costs and convergence to the fair price 
	is obtained assuming sensible parameters. 	
\end{abstract}	
	
\keywords{agent-based modeling, batch-mode reinforcement learning, market impacts, fitted Q-iteration, QLBS, FX option pricing, large trader}

\section{Introduction}

%*****

In our set-up, the existence of a large trader prevails in all sequences of the formulation of the problem. 
The large trader is defined as an entity in the environment of the reinforcement learning (RL) problem who has the capacity to affect transition from one state to another and aims to collect premium by issuing options. 
This explains the rationale behind our study. 
Therefore, we expect our contribution to converge to the fair price the large trader considers and at the same time deliver a more suitable and cost-efficient hedging strategy. 
Ultimately, optimal strategy leads to the implied exchange rate process the market (or the environment) observes. Through our reformulation we offer a testing environment, before issuing, for market participants that could alter the course of the markets. 
Our practical approach actually allows any sort of a large trader to foresee the possible effects of the transactions to be taken, and both visually and numerically delivers a testing environment. Basically, we reformulate our rationale as a pre-implementation wise.  

In contrast to the literature on market impacts as well as large trader models, that sides with the concept of no-arbitrage, our study is not built on the principle of no-arbitrage. 
Since our contribution is rather on reformulating the QLBS model, we leave a major characteristic of the model proposed by~\cite{halperin2020qlbs} intact. 
We therefore remain in touch with the data-driven nature of the study while extending into another dimension by including the existence of a large trader. 
Unlike~\cite{wilmott2000feedback}, we make no explicit assumption on a collection of small investors (i.e. agents). 
Rather we assume a large trader and an environment in which the trader sequentially moves. However, we also leave a room for collective behavior of such small investors.   
		
The trader issues a put option and devises a hedging strategy. 
However, in our approach, the exchange rate process is directly affected from the hedging strategy the trader postulates. 
Consider that the trader calculates a price to ask for which he considers fair. 
We assume that the large trader employs Monte Carlo simulation based on the trajectories affected by the hedging strategy he postulates. 
The trader first calculates the quoted exchange rate process based on the hedging strategy. 
We also address the issue of cost of repositioning, as each action requires an endurance of costs associated with both frequency and quantity the trader seeks. 

Given the embedded parameters of reformulation, the quoted exchange rate trajectories, in our experiments, might take unnatural courses. 
However, once the optimal hedging strategy is found, the implied exchange rate trajectories resembles more appropriate evolutions. 
This contributes to the fact that the  trader, whether be a trading desk of a large bank or an institutional client, might not come under fire for causing highly suspicious or disputable market practices. 
Another issue that we could encounter is resolved by assuming that the large trader is able to engage in any amount of transaction of the underlying, without the need of any execution of block by block or any sort for that matter.

We basically assume a feedback effect of a transaction. 
This effect is rather referred to as a \emph{feedback loop} in the terminology of RL, and this is why we can no longer employ the dynamic programming approach in generating optimal hedging strategy outlined in~\cite{dixon2020applications,halperin2020qlbs}. 
In~\cite{halperin2020qlbs}, this effect is ignored, and modeling is carried out under the standard assumptions of Black-Scholes-Merton model, 
i.e. no impact generated on the underlying. 
Though we employ the foreign exchange equivalent of BSM model, i.e. Garman-Kohlhagen~\cite{garman1983foreign} model, we do not follow standard assumptions. 
We, in the way provided by studies~\cite{roch2011liquidity,saito2017hedging}, relax the assumption of no impact generated on the underlying due to options prevailing  and allow the hedging activities to leave an impact on the evolution of the foreign exchange rate. 
Therefore, we contribute to the QLBS model by including the so-called feedback loop. 
Given the complexity related to the dynamics of alteration caused by the trader, we make several simplifications as well in regard to both the passage of time and the anticipation of market orders the large trader delivers. 
Besides, we assume no bid-ask spread for simplicity.   

We could state that our starting point jointly includes the literature on the large trader perspectives of financial markets and on market (i.e. price) impact models.
However, we lean further on the large trader aspects, rather than focusing on the order book's depth. 
There are several studies in regard to the large trader aspects we should mention, starting with~\cite{wilmott2000feedback} in which a clear distinction between the real value and the paper value, as delicately described by authors, is given. 
This approach is defined in terms of the illiquidity residing in the market. 
The replication problem of options under illiquidity is defined and nonlinear partial differential equation (PDE) for option replication is derived. 
Similarly,~\cite{frey1998perfect} could be consulted as well.
 
Though we keep volatility constant,~\cite{lions2007large} analyzes the possible effects of hedging by the large trader on the volatility. 
Similarly to our approach, only permanent impacts are considered. For a study solely focusing on temporary impacts,~\cite{rogers2010cost} should be considered. 
For an engaging description of temporary and permanent impacts,~\cite{almgren2001optimal} stands alone in which minimization of combinations of risk of volatility and costs of transaction are considered for both cases. 
Though we stand closer to studies in which the liquidity is introduced via an assumed supply curve, e.g.~\cite{cetin2006pricing,roch2011liquidity,saito2017hedging}, yet our approach also holds similarities to~\cite{bank2004hedging,jarrow1994derivative} in which the large trader constitutes an impact through a function of the amounts the trader holds.       

Moreover, the QLBS model, put forward by~\cite{halperin2020qlbs}, utilizes stock prices while we devise our reformulation for exchange rates and options written on exchange rates. 
We employ a finite-horizon risk-adjusted Markov Decision Process (MDP) in which the finite-horizon of the problem formulation is constrained to maturity of the contingent claim, same as the original study of the QLBS model. 
We do so by taking the finite-horizon the same as maturity, meaning that the life of the option ends in the formulated finite-horizon.

We extend such a flexible work on the grounds provided by~\cite{roch2011liquidity} and~\cite{saito2017hedging}. 
To the best of our knowledge, this is the first attempt in extending the QLBS model to include a large trader with a hypothetical limit order book and supply curve. 
In addition, we have not come across a similar way of handling on pricing options under market impacts with such flexibility, and we constitute this reformulation within the framework of MDP which is attempted on the works of neither~\cite{roch2011liquidity} nor~\cite{saito2017hedging}. 
Therefore, we also argue that our reformulation stands unique in the reinforcement literature as well, as our work is the first of its kind in bringing together RL, option pricing and market impacts given the discussion  we also provide in~\cite{ozsoy2023option}.

Another point we should emphasize is that the requirement of an optimal price along with optimal action values. 
We use action values and hedging strategy interchangeably, yet use the action values only for the solution provided by RL. 
In our illustrations, we show that as long as the input parameters are well adjusted for the model, in each case that optimal action values incur less cost in terms of transaction. 
Our reformulation, therefore, is stable, flexible and highly applicable. The choice of initial parameters and the simulation of the unaffected exchange rate process contribute to the flexibility of our proposed model. 
Another point is that the QLBS model is not built on the principle of ``no-arbitrage'', this also points out to the data-driven nature of the QLBS model. 
Therefore, in conjuncture with~\cite{halperin2020qlbs}, we put forward that the QLBS model under a large trader is not built on the principle of no-arbitrage.

\section{Formulation}

Equipped with a finite probability space $(\Omega, \mathcal{F}, \mathbb{P})$ along with fixed maturity $T$, due to how the reward function in the QLBS model is defined, 
our approach falls into the category of risk-sensitive Markov Decision Processes (hereafter, MDP or MDPs); 
for further inquiry on such MDPs, we refer to~\cite{gosavi2010finite}. 
The trader, as it is in discrete-time, takes a hedging decision sequentially at each time step, i.e. sequentially. To further elaborate, we define two exchange rate processes.

\begin{definition} \label{def_unaffected}
	Unaffected exchange rate process, $\{F_t\}_{t_0}^T \subseteq \R^+$, is the exchange rate process in which no hedging activities are taken.
\end{definition}  

\begin{definition} \label{def_quoted}
	Quoted exchange rate process, $\{\FF\}_{t_0}^T \subseteq \R^+$, is the exchange rate process that is directly affected by the hedging transaction postulated by the large trader at time $t$. 
\end{definition}

We also state that for this study  we assume no noticeable passage of time between $F_t$ and $\FF$, since in our study all changes related to hedging activities   \emph{instantaneously} take place, 
and happen in the exact same time step.

The trader is  modeled as a large trader whose actions in hedging transactions leave a rather \emph{lasting} impact on the quoted exchange rate. 
We therefore assume that there exists a large trader in the exchange rate market, that is sequentially acting under the optimal policy $\pi^*(\sS_t, t)$ given the state space, along with the actions $a=a(\sS_t,t) \in \mathbb{R}$. 
Since including the large trader through the framework MDP provides, with the actions expressed in terms of the state variable process denoted as $\{\sS_t\}_{t=0}^T$ at time $t$, 
we define time-dependent policy of the state variable $\sS_t$ and the time component $t$. 
Note that we denote the state variable explicitly by $\sS_t$ once we define unaffected exchange rate process and quoted exchange rate process. 

This rationale is the starting point as the large trader could be considered as the initiator of the quoted exchange rate process. 
Therefore, we aim to measure such impacts so that the large trader should be able to devise a better or a more efficient hedging plan given the impacts he single-handedly generates. 
We also assume that the hedging activities by the trader are so significant that the exchange rate dynamics are abrupt to some extent for which introducing two other parameters, impact and thinness, to be able to quantify the degree of the so-called impact on the exchange rate process becomes necessary. 

Moreover, we assume our trader, strictly, is a keen observer of the limit order book. 
We also assume in the light of the discussion of two parameters to be introduced. 
We assume the existence factors that leave an impact on the quoted exchange rate process; that is, the hedging activities of the trader, the market impact parameter, and thinness of the limit order book. 
We define one of the factors that alters the quoted exchange rate process as the market impact parameter; motivated by~\cite{roch2011liquidity}.

\begin{assumption} 
	There exists a exogenously given real-valued market impact parameter 
	$\beta_t$ at time $t$ such that $\beta_t \in [0,1)$. 
\end{assumption}

This parameter  leaves a permanent impact on the exchange rate given the hedging transactions through the limit order book, rather it is a measure of proportion of renewal. 
However, in~\cite{roch2011liquidity}, this parameter is taken constant. 
Yet we formulate our approach with a time-dependent market impact parameter, similar to that of~\cite{saito2017hedging}. 
We should mention that in case of $\beta_t \rightarrow 0$, there is  lower market impact for time $t$, however, if we set $\beta_t = 0$ then we assert  any possible market impact is neutralized, and lead to no alterations on the quoted exchange rate, for time $t$. 
Another parameter we shall introduce is the thinness of the order book.
A (hypothetical) limit order book is introduced via a supply curve. For a discussion on limit order books, we refer to~\cite{roch2013resilient} and references therein. In addition, examples of limit order books, and numerical illustrations, the study in~\cite{saito2015self} could be referred to. 
Therefore, following~\cite{roch2011liquidity,saito2017hedging}, we assume another factor contributing the alteration of the quoted exchange rate process. 

\begin{assumption}\label{prop_3} % taken as same of the thesis
	There exists    $\{M_t\}_{t=0}^{T} \subseteq \R^+$ which refers to the thinness of the order book, also defined as a measure of illiquidity by~\cite{roch2011liquidity}.
\end{assumption}

Therefore, we conclude the factors that affect the quoted exchange rate process, and move to the specifics of the supply curve that the trader is able to observe in terms of the signal he receives via the environment. 
Our horizon is finite and bounded by $T$ and $\FF$ is the quoted exchange rate. Furthermore, assume $u_t$ refers to the hedge position of the underlying at time $t$. 
Then, the unaffected exchange rate process ${F}_t$ is an adapted continuous process, and the unaffected exchange rate dynamics is determined exogenously. 
Following~\cite{roch2011liquidity} and motivated by~\cite{ccetin2004liquidity}, the unaffected supply curve is therefore given by
	\begin{equation} \label{marketimpact:linearcurve}
		F_t (u_t) = F_t + M_t u_t  
	\end{equation}
for $t= 0, \ldots, T$.
An implication of \eqref{marketimpact:linearcurve} is that the assumed supply curve corresponds to the linear structure per se. 
As also pointed out in~\cite{roch2011liquidity}, this form is empirically supported by the study~\cite{blais2010analysis}. 
An empirically profound study,~\cite{blais2010analysis}, shows that supply curves exist as proposed by~\cite{ccetin2004liquidity}, and the assumed structure is not rejected for actively traded stocks that are being traded in greater volumes and depth. 
Given the conclusions in the above-mentioned study, we assume such a structure is also applicable in foreign exchange markets. However; we, to the best of our knowledge, are unaware of such studies repeated in terms of currencies. 

One word of caution regarding \eqref{marketimpact:linearcurve}  from ~\cite{roch2011liquidity} is that there could be a possibility indicating that exchange rate process might take negative values; 
however, as long as $M$ is sufficiently small such likelihoods are unlikely. 
Following~\cite{ccetin2004liquidity,roch2011liquidity}, we impose observability of the unaffected exchange rate process $\{F_t\}_{t=0}^{T}$ by assuming that supply curve of the unaffected exchange rate is assumed to be adapted to the filtration $\mathcal{F}$ until the horizon of the formulated learning problem.

\subsection{Hedge Portfolio}

The unaffected exchange rate process, $\{F_t \}_{t=0}^{T} \subseteq \R^+$, defined as time-indexed spot price of the currency to be delivered at maturity, is taken, throughout our reformulation, in terms of domestic units per foreign units, see~\cite{garman1983foreign}. 
Constant domestic and foreign interest rates such as $r^\text{d}, r^\text{f} \in \R^+$ are assumed.
In order to liquidate the hedge portfolio at expiration, we set the hedge position at maturity to zero, with the postulated hedge strategy $\{u_t\}_{t=0}^T \subseteq \R$. 

Similar to the original QLBS model, we approach to the pricing problem from the perspective of an issuer; this is the large trader in our case. 
However, an important clarification is needed: we assume that the contingent claim is cash-settled and the difference between the strike and the exchange rate at expiration is paid if the option ends in-the-money. 

Consider a European  put option written on foreign exchange rate (i.e. currency options) the trader is planning on issuing, we assert that there exists a bounded maturity $T^+\in \R^+$ and the payoff $\delta(F_{T^+}) = \max(K-F_{T^+},0)$ of the contingent claim at maturity.

A hedge portfolio $\Pi_t$ is formed with the underlying asset, and a risk-free asset denominated in the domestic currency, i.e. money market account.
Moreover; $\Pi_{T^+} = B_{T^+} = \delta(F_{T^+})$, i.e., a terminal condition is levied. Motivated by~\cite{grau2008applications} and~\cite{halperin2020qlbs}, we propose the hedge portfolio associated to the hedging strategy recursively given as follows:
\begin{equation} \label{Eq A26}
		\Pi_t = e^{-r^\text{d} \Delta t } [\Pi_{t+1} - u_t \Delta \FF     ] , ~~~~\Delta \FF = \FFFF - e^{r^\text{d} \Delta t} \FF %, ~~~~  t = T-1, \ldots, 0.
\end{equation} 
for $t = T-1, \ldots, 0$.

In contrast to \eqref{Eq A26}, both processes $\Pi_t$ and $B_t$ are observable at option's maturity. 
The reason we emphasize the construction of the hedge portfolio is that, in the QLBS model, selection of the hedge strategy does not alter the dynamics of underlying. 
The QLBS model stays within the classical assumptions of the Black-Scholes-Merton model; yet, in our proposed model it does. 
Every possible hedging strategy postulated by the large trader generates a unique trajectory of the quoted exchange rate process. 
This is rather the very essence we propose in our reformulation of the QLBS model, and main contribution to the related literature.

\subsection{Quoted Exchange Rate Dynamics}

Here we discuss further specifics of the quoted exchange rate process. Since we aim to quantify how trajectories of the quoted exchange rate process, we need to define the specifics of such processes. 
As we integrate the market impact perspective into the QLBS model, we should mention first that we are greatly inspired by both~\cite{roch2011liquidity} and~\cite{saito2017hedging}, and build upon their contributions to the literature. 
By \eqref{marketimpact:linearcurve}, we  construct the (hypothetical) limit order book, and define $\rho_t(z)$ as the density and the amount of available underlying  currency at rates $z$ between $z_t$ and $z_{t+1}$ is given by 
\begin{equation}\label{eq:density}
	\int_{z_t}^{z_{t+1}} \rho_t(z)dz. 
\end{equation}
Therefore, 
when the large trader is to take a position $u_t$, the cost associated is given by $\int_{F_t}^{z_{u}} z \rho_t(z)dz$, with  $z_{u}$ solving $\int_{F_t}^{z_u} \rho_t (z)dz$.
Notice that the supply curve, represents the limit order book of all the  orders of the market participants. 
As also pointed out by~\cite{roch2009liquidity}, it is so hypothetical that is  unobservable directly, yet given this structure we are able to obtain the supply curve of the quoted exchange rate process. 
Given the formulation of the supply curve, 
the density is $\rho_t = \frac{1}{2M_t}$, which yields $z_u= F_t +2M_tu$, 
for more  details we refer to ~\cite{roch2011liquidity}. 

We point a direct result of the above discussion, and of the specifications of the unaffected supply curve. 
Consider the associated transaction cost of $u$ amount of the underlying, i.e.;
\begin{equation}
	\frac{1}{2M_t} \int_{F_t}^{F_t + 2 M_t u} z dz = F_t u+ M_t u^2,
\end{equation}
in which we could spot that greater values of $M_t \in \R^+$ lead to greater costs incur. This actually explains why it is defined as a measure of illiquidity as well.

Let us explain the general case, as laid out in~\cite{roch2011liquidity}, of exchange rate dynamics given the hedge position the large trader takes. 
We also now consider the effect of such transactions on the limit order book. 
Consider an arbitrary market order to buy $u_t$ amount of the underlying, given the difference between two consecutive repositioning denoted by $\Delta u_t$. 
The large trader is to add $\Delta u_t$ amount of the underlying currency to his hedge portfolio; this leads to the new quoted exchange rate, $F_t + 2 \beta_t M_t \Delta u_t$. 
The interpretation of  $\beta_t $ is to consider $1-\beta_t$ as the fraction of the limit order book which is to be renewed upon a transaction the large trader engages. 
Following the analogy of~\cite{saito2017hedging}, we let $F_t^0$ denote  mid-price of the exchange rate process, the quoted exchange rate process could be referred to as the average price. 

It is instinctive to strike a question on whether or not the so-called large trader is able to take positions as he sees fit. 
This question, rather, nurtures the concept of timing which could be an unavoidable matter in further formulations aimed at capturing rather complex market dynamics. 
However, in our sole large trader approach, we assume the trader is able to engage in transactions regardless of the hedging positions. 
This, indeed, explains our reason to remark that the exchange rate market anticipates and any amount of the underlying could be made available given the thinness of the order book. 
Motivated by~\cite{roch2009liquidity,roch2011liquidity,saito2017hedging}, we specify the mid-price  associated to the self-financing hedging strategy \eqref{Eq A26} for $t = T, \ldots, 0$, and $\Delta u_0=u_0$ as
\begin{equation} \label{eq:marketimpact*}
		F_t^0 =  F_t + \sum_{j=0}^{t-1} 2 \beta_{j } M_j \Delta u_j.
\end{equation}
Hence, by \eqref{eq:marketimpact*} we quantify the quoted exchange rate process, $\{\FF\}_{t=0}^{T}$, which is associated to the self-financing hedging strategy postulated by the large trader in \eqref{Eq A26}. 
Also, we assume the large trader pays $\FF \Delta u_t$ from $t-1$ to $t$, and by~\cite{roch2009liquidity,roch2011liquidity,saito2017hedging} and  \eqref{eq:marketimpact*}, we put forward that the quoted exchange rate process %$\{\FF\}_{t=0}^T \subseteq \R^+$
evolves through the following dynamics
\begin{equation}
		\FF = F_t^0 + 2 \beta_t M_t \Delta u_t.
\end{equation}
Thus, we now impose observability condition on the quoted exchange rate, $\{\FF\}_{t=0}^T$, as well such that the supply curve of the quoted exchange rate is assumed to be adapted to the filtration $\mathcal{F}_{t^+}$ until the horizon of the learning problem  formulated. 
 
The large trader is able to estimate the unaffected exchange rate process, $\{F_t\}_{t=0}^T$, within a hypothetical boundary of error rate. 
We, for this study, prefer not to specify the aforementioned error rate, the preferences, beliefs and forecasting capability of the trader are assumed to be included. 
Methodology is as follows; given the past behavior and forward-looking evolution of the exchange rate process, $\{F_t\}_{t=0}^T$, the trader plans his hedging activities; namely, $F_0 \rightarrow F_1 \rightarrow F_2 \rightarrow \cdots \rightarrow F_T$, and $u_T=0 \rightarrow u_{T-1} \rightarrow u_{T-2} \rightarrow \cdots \rightarrow u_0$ are formulated.

We remark that the predetermined hedging strategy, $\{u_t\}_{t=0}^T$, need not be optimal given the unaffected exchange rate process. 
We require no necessities. However, a major reason we assume non-necessity is the fact that Q-learning is all on off-policy. 
However, we are aware of the fact that there are budget constraints in reality, which we ignore in this paper. 
Nevertheless, we indicate that non-optimality of the predetermined hedging strategy, $\{u_t\}_{t=0}^T$, is acceptable. 
We reemphasize that the large trader  is faced with a supply curve that is affected by the size of his orders.
 
The large trader \emph{postulates} the quoted exchange rate process for which he assumes himself a driving force. 
To what extent the unaffected exchange rate process moves under the effects of his transactions is a motivating question from his perspective. 
Especially, the difference between the hedging strategy  he postulates and the optimal hedging strategy corresponding to the referred hedging strategy matters, not only in terms of the costs of hedging activities but also from the perspective of the fair price to be asked. 
Although we have no reason to make things complicated, however, the large trader already determines a price to charge. 
In that, our question is that whether the price generated by the proposed extended QLBS model could converge to the perceived fair price while generating significantly lesser cost. 

Though other approaches could be consulted, in this paper, the large trader calculates the fair price through plain Monte Carlo simulation through the quoted exchange rate process, not through the unaffected exchange rate process. 
Let $\mathcal{C}^{\text{fair}}_t$ denote the fair put option price from the perspective of the large trader at time $t$, and recall we assume the passage of time between $T$ and $T^+$ is so negligible that we only utilize such a difference for notational purposes:
\begin{equation}
 \mathcal{C}^{\text{fair}}_0 = e^{-r^\text{d} T} \frac{1}{N_{\text{MC}}} \sum_{j=1}^{N_{\text{MC}}} \max\{ K- F_{T^+}^j,0 \},
\end{equation}
where $N_{\text{MC}}$ is the number of replications (paths) used in Monte Carlo simulations. Given the unaffected supply curve, and the corresponding unaffected exchange rate process, 
that is $F_0 \rightarrow F_1 \rightarrow F_2 \rightarrow \cdots \rightarrow F_T$, a predetermined hedging strategy $\{u_t\}_{t=0}^T$ is devised. 
Further, the quoted exchange rate process is postulated given a  predetermined hedging strategy $\{u_t\}_{t=0}^T$, that yields $\{ \FF(u_t) \}_{t=0}^T$, indicating the quoted exchange rate process depends on the order size. 
%Given a fixed boundary $t<T$, 
Hence, the sequence $F_{0^+}(u_0) \rightarrow F_{1^+}(u_1) \rightarrow  F_{2^+}(u_2) \rightarrow \cdots \rightarrow  F_{T^+}(u_T=0)$ yields the optimal hedge strategy and the optimal price to ask.

We now define another exchange rate, namely, the implied exchange rate process 
which corresponds to the exchange rate process generated through the optimal actions. 
Since the postulated hedging strategy is related to the quoted exchange rate process, the implied exchange rate process is generated by optimal action values once the quoted exchange rate along with the postulated hedge strategy is utilized.
\begin{definition}\label{imp_FX}
	Implied exchange rate process is the process 
	 $\tilde{F}_{t^+}(a_t^*)$ for $t=0,\ldots,T$
	formulated exactly the same as quoted exchange rate process, yet include the optimal action values as the hedging strategy, rather than the postulated hedge strategy values.
\end{definition}
Hence, this definition implies $\tilde{F}_{0^+}(a_0^*) \rightarrow \tilde{F}_{1^+}(a_1^*) \rightarrow  \tilde{F}_{2^+ }(a_2^*) \rightarrow \cdots \rightarrow \tilde{ F}_{T^+}(a_T^*=0)$.
We now propose that the large trader wishes to collect a fair price (premium) based on the quoted exchange rate process such that Monte Carlo prices based on the quoted exchange rate process and the implied exchange rate process converge. 
In the mean time, they generate an optimal hedging strategy that yields lower cost of transaction costs,
along with the QLBS price to ask based on the implied exchange rate process. 
In essence, we feed the algorithm with the postulated hedging strategy, and the corresponding quoted exchange rate processes. 
In return, we aim to find a price that is expected to converge to the fair price, and optimal hedge strategy. 
Ultimately, we aim to result in the implied exchange rate process that is generated by the optimal action values. 

For a discount factor $0 \le \gamma \le 1$, assume the agent transitions from one state to another at time $t$ with transition probability $ p(s',r \mid s, a)$ given $\mathcal{S}_t \in \mathcal{S}$ by taking  actions such as $a_t \in \mathcal{A}$ with $\mathcal{A}$ being the set of actions. The agent receives rewards, $\mathcal{R}_t \in \mathcal{R}$, at each state $\mathcal{S}_t$. Therefore, defining the necessary elements of a Markov Decision Process, we have a tuple such that $\{ \mathcal{S}, \mathcal{A}, p(s',r \mid s, a), \mathcal{R}, \gamma \}$. 

We now further contribute to the QLBS model by providing a more flexible and practical way of constructing the state variables process, we assume the following state-space.

\begin{assumption}\label{log_impact}
	For $t= 0,\ldots,T$, we assume the dynamics %%and $\{\mathcal{S}_{t}\}_{t=0}^T \in \R$, $\{\FF\}_{t=0}^T \in \R^+$
	%\quad
	\begin{equation}\label{log_ret}
		\mathcal{S}_{t}  =  \ln \left(\frac{\FF }{\FFF}\right), \quad \{\mathcal{S}_{t}\}_{t=0}^T \subseteq \R, \quad \{\FF\}_{t=0}^T \subseteq \R^+,
	\end{equation}
	provided that $\FFF \neq 0$.
\end{assumption} 

For discussion and numerical tests on applicability of the approach based on the QLBS given in Assumption \ref{log_impact}, we refer to~\cite{ozsoy2023option}. 

We now develop further specifics. Let us return to the fair price concept of the QLBS model for which the trader should charge upon issuing. 
We argue the trader  cannot ask for the fair price, and adding a compensation term. 
Following~\cite{halperin2020qlbs}, we assume the risk aversion parameter $\lambda$ which is a positive constant that measures the trader's preference of risk.
% such as $\lambda \in \R^+$. 
Following~\cite{halperin2020qlbs}, the European currency option price is given by
%	For $t= 0,\ldots,T$, and $\lambda > 0$
\begin{equation} \label{eq:C_0_impact}
		C_0^{\text{ask}}  = \mathbb{E}_0 \left[  \Pi_0 + \lambda \sum_{t=0}^{T}  e^{-r^\text{d} t} \mathbb{V}\text{ar}[ \Pi_t \mid \mathcal{F}_{t^+} ] \middle| F_{0^+}, u_0      \right].
\end{equation}
We should note that in \eqref{eq:C_0_impact} that we condition upon $F_{0^+}$ and $u_0 $. 
We revise  \eqref{eq:C_0_impact} as a maximization problem, i.e. $V_t =-C_0^{\text{ask}}$, that yields the value function
\begin{equation}\label{imp_value_max}
	V_t (\sS_t) = \mathbb{E} \left[  -\Pi_t - \lambda \sum_{t^{'} =t}^{T} e^{-r^\text{d}(t^{'}-t)} \mathbb{V}\text{ar} [\Pi_{t^{'}} \mid \mathcal{F}_{t^{'}}] \middle| \FF   \right],
\end{equation}
We also note that adding a premium in proportion to the variance of the self-financing hedging strategy is first suggested by~\cite{potters2001hedged},
as also called out by~\cite{halperin2020qlbs} in the original QLBS model study.

\subsection{Value Function and Rewards}

We assume that the large trader observes the hypothetical limit order book of the underlying and is allowed to engage in any amount of transaction as we avoid imposing such constraints. 
Given the state variables $\{\sS_t \}_{t=0}^T$, we express the action variables, $ u_t (\FF ) = a_t (\sS_t(\FF))$, and referring to the actions taken in a particular state $\sS_t \in \sS $. 
We, given the formulation of the QLBS model in~\cite{halperin2019qlbs,halperin2020qlbs}, utilize the rewards as negatives of the hedge portfolio one-step variances times the risk aversion along with a drift term.  

We present the notion of deterministic time-dependent policy function. 
Policy $\pi(\sS_t,t)$ determines the hedging actions, i.e. $a_t = \pi (t, \sS_t)$, and we are in a Batch-mode reinforcement learning. 
Motivated by~\cite{halperin2020qlbs}, the value-maximization problem laid in we propose  the value function as
\begin{align} \label{eq:impackeq7}
		V_t^{\pi} (\sS_t) &= \mathbb{E}_t \left[ -\Pi_t(\sS_t)  - \lambda \sum_{t^{'} =t}^{T} e^{-r^\text{d}(t^{'}-t)} \mathbb{V}\text{ar} [\Pi_{t^{'}} \mid \mathcal{F}_{t^{'}}] \middle| \FF   \right],  \\
		&= \mathbb{E}_t \left[ -\Pi_t(\sS_t) -\lambda \mathbb{V}\text{ar} [\Pi_t] - \lambda \sum_{t^{'} =t+1}^{T} e^{-r^\text{d}(t^{'}-t)} \mathbb{V}\text{ar} [\Pi_{t^{'}} \mid \mathcal{F}_{t^{'}}] \middle| \FF  \right]. \nonumber
\end{align}
The upper index, denoted by $\pi$  in  \eqref{eq:impackeq7}, indicates the dependence of the value function on the deterministic policy that maps time and the state variable $\sS_t$ into action taken, $a_t \in \mathcal{A}$, at that particular state. 
%Given \eqref{eq:impackeq7}, 
We rearrange the terms explicitly and in terms of the next-step value function, and obtain
\begin{equation}\label{eq3_marketimpact1}
	-\lambda \mathbb{E}_{t+1} \left[   \sum_{t^{'} =t+1}^{T} e^{-r^\text{d}(t^{'}-t)} \mathbb{V}\text{ar} [\Pi_{t^{'}} ]   \right] =\gamma (V_{t+1} + \mathbb{E}_{t+1 } [\Pi_{t+1}]), %~~~ \gamma = e^{-r^\text{d} \Delta t}.
\end{equation}
where $\gamma = e^{-r^\text{d} \Delta t}$.
Using \eqref{eq:impackeq7} and \eqref{eq3_marketimpact1}, and motivated by~\cite{halperin2020qlbs}, we obtain the following Bellman equation for the proposed extended QLBS model; for $t=0,\ldots,T-1$, %and 	$\gamma = e^{-r^\text{d} \Delta t}$
\begin{equation} \label{marketimpact_eq5}
		V_t^{\pi}(\sS_t) = \mathbb{E}_t^{\pi} [\mathcal{R}(\sS_t,a_t,\sS_{t+1})        +\gamma V_{t+1}^{\pi} (\sS_{t+1})   ].
\end{equation}
We  should draw attention to how consecutively tied two states are, since the large trader is formulated as a sequential decision maker. 
Another critical component of MDPs is the existence of rewards. 
Given a collection of rewards $\mathcal{R}$; in each state there exists a random reward %such as 
$\mathcal{R}_t \in \mathcal{R}$ for $t=0,\ldots,T$. 
Motivated by~\cite{gosavi2010finite,halperin2020qlbs}, one-step time-dependent random rewards are given by  
\begin{align}\label{eq5_marketimpact1}
		\mathcal{R}(\sS_t,a_t,\sS_{t+1}) &= \gamma a_t \Delta \FF - \lambda \mathbb{V}\text{ar}[\Pi_t \mid \mathcal{F}_{t^+}], \notag \\
		&= \gamma a_t \Delta \FF - \lambda \gamma^2 \mathbb{E}_t\left[  \hat{\Pi}_{t+1}^2 -2a_t \Delta \hat{F}_{t^+} \hat{\Pi}_{t+1}+ a_t^2 (\Delta \hat{F}_{t^+} )^2      \right],
\end{align}
in which $\Delta \hat{F}_{t^+} = \Delta \hat{F}_{t^+}- {\Delta \bar{F}_{t^+}}$ is utilized. 
In \eqref{eq5_marketimpact1}, we used $\hat{\Pi}_{t+1} = \Pi_{t+1}- \bar{\Pi}_{t+1} $ where 
$\bar{\Pi}_{t+1}$ is the sample mean of all  values of $\Pi_{t+1}$.
% and this holds for $\Delta \FF$ as well. 
Note also that for $t=T$, we have $\mathcal{R}_T = - \lambda \mathbb{V}\text{ar} [\Pi_T]$.

\subsection{Action-value Function}

We define the  corresponding action-value function, or Q-function, given the quoted exchange rate process, and actions, respectively, $\{\FF\}_{t=0}^T \subseteq \R^+$ and $\{a_t\}_{t=0}^T \subseteq \R$. 
Action-value functions are defined and conditioned upon both the state variable and the initial action. 
In fact, Q-function is obtained by taking the expectation of \eqref{eq:impackeq7}. Given $\sS_t \in \sS$ and $a_t \in \mathcal{A}$, it follows that
\begin{multline} \label{act3_1}
		Q_t^{\pi} (s,a) =\mathbb{E} [-\Pi_t(\sS_t) \,|\, s, a    ]  
		-\lambda \mathbb{E}_t^{\pi} \left[\sum_{t^{'} =t+1}^{T} e^{-r^\text{d}(t^{'}-t)} \mathbb{V}\text{ar} [\Pi_{t^{'}} \mid \mathcal{F}_{t^{'}}] \middle| s, a \right].
\end{multline}
In \eqref{act3_1},  the second term is dependent on the policy as the first term merely averages the next time-step. 
The Bellman equation of the Q-function of our reformulation  under market impacts is obtained similarly to that of the derivation of \eqref{marketimpact_eq5}:
\begin{equation}
		Q_t^{\pi}(s,a) = \E_t [ \mathcal{R}(\sS_t,a_t,\sS_{t+1}  )\mid \sS_t = s, a_t=a     ] + \gamma \E_t^{\pi} [V_{t+1}^{\pi}(\sS_{t+1}) \mid \sS_t=s].
\end{equation}
The optimal policy $\pi^*$ is defined as the policy that maximizes  the value-function or alternatively the action-value function:
\begin{equation}
	\pi_t^{*}(X_t) = \arg \max \limits_{\pi} V_t^{\pi} (X_t) = \arg \max \limits_{a_t \in \mathcal{A}} Q_t^{\pi} (x,a).
\end{equation} 

We remark that the Bellman optimality equation for  action-value function with a terminal condition 
	\begin{multline} \label{eq6_1_ext}
		Q_t^{\pi} (s,a) =\mathbb{E}_t^{\pi} [\mathcal{R}(\sS_t,a_t,\sS_{t+1})        
		\\ +\gamma \max \limits_{a_{t+1} \in \mathcal{A} } Q_{t+1}^{\pi} (\sS_{t+1},a_{t+1}) \mid  \sS_t = s, a_t = a  ], %\\
	\end{multline}
	with a terminal condition  set at $t=T$ given by
	\begin{equation}
		Q_T^{\pi} (\sS_T,a_T=0) = \Pi_T(\sS_T) - \lambda \mathbb{V}\text{ar}[\Pi_T(\sS_t)]
	\end{equation}
at $t = T$.
In \eqref{eq6_1_ext}, the terminal value of hedge portfolio is calculated by the specified terminal condition and variance is calculated based on all (possibly, Monte Carlo) paths that terminate in a given state. Explicitly, for $t=0,\ldots, T-1$,
	\begin{multline}\label{bellman1_ext}
		Q_t^{*} (\sS_t,a_t) = \gamma \mathbb{E}_t \left[   Q_{t+1}^{*} (\sS_{t+1},a_{t+1}^*) + a_t \Delta \FF        \right] \\
		- \lambda \gamma^2 \mathbb{E}_t \left[  \hat{\Pi}_{t+1}^2 -2a_t \Delta \hat{F}_{t^+} \hat{\Pi}_{t+1}+ a_t^2 (\Delta \hat{F}_{t^+}^2) \right],
	\end{multline}
for which we substitute the expected reward function into the Bellman optimality. 
Notice that if we were to have an explicit expression for the optimal values, \eqref{bellman1_ext} would also yield an optimal action-value function.

\section{Fitted Q-iteration for the Extended Model }

The QLBS model is developed under the assumptions of the Black-Scholes-Merton model, i.e. no impact on the underlying. 
Therefore, we reformulate and readjust the rationale behind our reasoning to be able to price currency options under market impacts. 
The methodology that appears in the QLBS model could also be used for our case. 
However, in contrast to the original model, we do now have a dynamic programming approach as there is a dependency between consecutive actions. 
We argue that we can no longer employ backward recursion for the Q-function. 
Besides, we do not provide our trader with optimal or sub-optimal action values either. 

We are in the batch-mode learning, assuming we have a collection of data that is applicable for our motivation. 
A highly important aspect of batch-mode reinforcement learning is that there is no need to know transition probabilities or reward functions. 
This amounts to the very reason that this approach appeals to our reformulation.  

Methodology is to first, given the past behavior and forward-looking evolution of the exchange rate process, $\{F_t\}_{t=0}^T$, the large trader devises his hedging activities. 
That is, $F_0 \rightarrow F_1 \rightarrow F_2 \rightarrow \cdots \rightarrow F_T$, and $u_T=0 \rightarrow u_{T-1} \rightarrow u_{T-2} \rightarrow \cdots \rightarrow u_0$ are formulated. 
In the original QLBS model, as discussed, the predetermined hedging strategy does not cause alterations; 
however, in our case, each set of the predetermined hedging strategy, whether it is optimal or non-optimal, causes alterations in the evolution of the quoted exchange rate process $\{\FF\}_{t=0}^T$ from the forward-looking sense. 
A direct result of the approach eliminates dynamic optimization and leaves us to the realm of reinforcement learning. 
Since the batch-mode learning requires the state-space and each predetermined hedging strategy yields a state space different from the assumed underlying exchange rate process, the algorithm is fed with the state space stemming from the predetermined hedging strategy devised. 

Given the unaffected supply curve, and the corresponding unaffected exchange rate process the large trader hypothetically observes up to time of initialization, and   $F_0 \rightarrow F_1 \rightarrow F_2 \rightarrow \cdots \rightarrow F_T$, a predetermined hedging strategy is devised, $\{u_t\}_{t=0}^T$. 
Further, the quoted exchange rate process is \emph{postulated} given a   predetermined hedging strategy $\{u_t\}_{t=0}^T$, that yields $\{ \FF(u_t) \}_{t=0}^T$, indicating the trajectory of the quoted exchange rate process depends on the order size along with the market impact parameter, $\beta_t$, and the thinness of the order book, $M_t$. 
Given a fixed boundary $T$, $F_{0^+}(u_0) \rightarrow F_{1^+}(u_1) \rightarrow  F_{2^+}(u_2) \rightarrow \cdots \rightarrow  F_{T^+}(u_T=0)$ yields the optimal hedge strategy and the optimal option price to ask. 
This procedure could be repeated as many times as required. By requirement, we could include the large trader's acceptable hedging expenses or hedging error rate. 
This possibility of starting with a predetermined hedge strategy and postulating a trajectory of the quoted exchange rate process increases the flexibility of the  contribution of the paper. 
Indeed, this could also be used in the perspective of risk management as it is planned (postulated) before issuing.

In this setting, $N_{\text{MC}}$ denoted as the number of paths of a bounded length $T$ yields a data set of $N_{\text{MC}} \times T$ given a predetermined hedging strategy $\{a_t^{(n)}\}_{t=0}^T$ devised for each of the paths (or trajectories).
For this we follow~\cite{halperin2019qlbs,halperin2020qlbs} in revising for our extension.
We implement Fitted Q-iteration,~\cite{ernst2005tree,murphy2005generalization}. 
In this approach, all Monte Carlo paths of the quoted exchange rate are utilized. 
Taking path-wise empirical means of all the quoted exchange rate processes is, basically, the essence. 
The set-up is now based on inputs, namely samples of different variables provided for the algorithm. 
The only data available is given by a set of $N_{\text{MC}}$ paths for the underlying state variable $\sS_t$, hedge position $a_t$, instantaneous reward $\mathcal{R}_t$, market impact $\beta_t$, thinness of the order book $M_t$, and the next-step  $\sS_{t+1}$. 
Thus, we assume the following structure of the sample 
% is available, for $ k=1,\ldots,N_{\text{MC}}$
%
\begin{equation}\label{data_av}
		\mathcal{F}_t^k=\left\{\left(\sS_t^k,\sS_{t+1}^k,a_t^k,\mathcal{R}_t^k,\beta_t^k,M_t^k\right)\right\}_{t=0}^{T-1}
\end{equation}
for $ k=1,\ldots,N_{\text{MC}}$.
Notice that in \eqref{data_av}, the quoted exchange rate process $\{\FF^k\}_{t=0}^T$ is not directly incorporated, 
yet embedded through the state variables that is computed by \eqref{log_ret}. 
The reality of our proposed model might be seen restrictive in this case, 
however, the large trader only postulates assuming he is to shift the exchange rate given his size of transactions. 
This would require the entire path of the quoted exchange rate fixed at some values of $k$, making the change of variables given in~\cite{halperin2020qlbs} unusable in this set-up. 

We utilize a set of basis functions $\mathbf{\{\Phi_{n}(s)\}}$. 
Our selection coincides with studies in~\cite{halperin2019qlbs,dixon2020applications}, as in  both 12 basis functions chosen to be (cubic) B-splines that is with a range from smallest to greatest values observable in data. 
The optimal Q-function $Q^*_t(\sS_t, a_t)$ could be represented as an expansion in the basis functions. Given weights matrix $\mathbf W_t$ and $\mathbf \Phi_{n}(s)$ as
\begin{equation}
	\mathbf W_t  = \begin{bmatrix}
		W_{11}(t) &W_{12}(t) &\ldots& W_{1M}(t) \\ 
		W_{21}(t)& W_{22}(t)& \ldots &W_{2M}(t) \\
		W_{31}(t)& W_{32}(t)& \ldots &W_{3M}(t)
	\end{bmatrix},
\mathbf{	\Phi_{n}(s)} =  \begin{bmatrix}
		\Phi_1(s) \\ \vdots \\ \Phi_M(s)
	\end{bmatrix},
\end{equation}
by~\cite{halperin2020qlbs}, the optimal Q-function can be  expressed as 
\begin{equation}\label{rev1}
		Q_t^\star\left(\sS_t,a_t\right)=\mathbf A_t^\top \mathbf{W}_t\mathbf{\Phi}\left(\sS_t\right)
		= \mathbf{A}_t^\top \mathbf{U}_W\left(t,\sS_t\right) = \vec{ \bf W}_t^\top \vec{\Psi}\left(\sS_t,a_t\right) %=\vec{W}_t^\top \vec{\Psi}\left(\sS_t,a_t\right),
\end{equation}
for given $\sS_t \in \sS$ and $a_t \in \mathcal{A}$, where
\begin{equation}
		\mathbf A_t= \begin{bmatrix}1\\a_t\\\frac{1}{2}a_t^2\end{bmatrix}, 
		\quad \mathbf{U}_W\left(t,\sS_t\right)=\mathbf {W_t\Phi(\sS_t)} \notag.
\end{equation}
We rearrange \eqref{rev1} as
\begin{align}\label{ham_knoc}
	Q^*_t(\sS_t,a_t) &=\mathbf{A}_t^\top \mathbf{U}_W\left(t,\sS_t\right) = \sum_{i=1}^{3} \sum_{j=1}^{M} \left(  \mathbf{W}_t \odot \left( \mathbf{A}_t \otimes \mathbf{\Phi}^\top(\sS)     \right)    \right)_{ij} \\
	&= \vec{\mathbf{W}}_t \cdot vec \left( \mathbf{A}_t \otimes \mathbf{\Phi}^\top (\sS)    \right) = \vec{\mathbf{W}}_t \vec{{\Psi}}(\sS_t,a_t), \nonumber
\end{align}
where
$\odot$ refers to element wise (Hadamard) product of matrices, whereas $\otimes$ refers to outer (Kronecker) product of matrices. 
The vector $\vec{\mathbf{W}}_t$ is obtained by concatenating columns of matrix $\mathbf W_t$ while $   \vec{{ \Psi}} \left(\sS_t,a_t \right)  = vec \, \left( {\bf A}_t  \otimes {\bf \Phi}^\top(\sS) \right) $ refers to a vector obtained by concatenating columns of the outer product of vectors $ {\bf A}_t $ and $ {\bf \Phi}^\top(\sS) $. 

We are in need of coefficients of $\mathbf{W}_t$ that could be computed in a recursive manner, i.e. $t= T-1, \ldots, 0$. 
Recall the equation of Bellman optimality in \eqref{eq6_1_ext}, which is to be presented as a regression with the following form, with $\epsilon_t$ defined as a random noise with zero mean at $t$: %that is
\begin{equation}\label{reg1}
	\mathcal{R}_t(\sS_t, a_t , \sS_{t+1}) + \gamma \max \limits_{a_{t+1} \in \mathcal{A} } Q^*( \sS_{t+1}, a_{t+1}) = \vec{\mathbf{W}}_t \vec{\Psi{}}(\sS_t,a_t) + \epsilon_t.
\end{equation}
This leads to the following least-squares optimization of the objective:
\begin{equation}\label{min1_*}
\mathcal{L}_t (\mathbf{W}_t) = \sum_{k=1}^{N_{\text{MC}}}\left(	\mathcal{R}_t(\sS_t, a_t , \sS_{t+1}) + \gamma \max \limits_{a_{t+1} \in \mathcal{A} } Q^*( \sS_{t+1}, a_{t+1}) - \vec{\mathbf{W}}_t \vec{{\Psi}}(\sS_t,a_t)\right)^2.
\end{equation}
As put forward by~\cite{halperin2020qlbs}, \eqref{min1_*} holds only for the setting of the off-policy of the Fitted Q Iteration.
We first compute vector $\mathbf A_t$, then compute $\vec\Psi\left(\sS_t,a_t\right)$ for each $\sS_t^k$. The vector $\vec{\mathbf{W}}_t$ is therefore solved  by the solution of $\mathbf S_t \vec{\mathbf{W}}_t^*= \mathbf M_t$ leading to a matrix $\mathbf{S}$ of size $n \times m$ and a vector ${\bf M}$
\begin{align}\label{set2}
	{S}_{nm}^{\left(t\right)}&=\sum_{k=1}^{N_{\text{MC}}}{\Psi_n\left(\sS_t^k,a_t^k\right)\Psi_m\left(\sS_t^k,a_t^k\right)}, \\ {M}_n^{\left(t\right)}&=\sum_{k=1}^{N_{\text{MC}}}{\Psi_n\left(\sS_t^k,a_t^k\right)\left(\mathcal{R}_t\left(\sS_t^k,a_t^k,\sS_{t+1}^k\right)+\gamma\max_{a_{t+1}\in\mathcal{A}}Q_{t+1}^\star\left(\sS_{t+1}^k,a_{t+1}\right)\right)}.
\end{align}
The proof of \eqref{set2} is straight-forward. Differentiating \eqref{min1_*} with respect to vector $\vec{ \bf W}$ and setting the equation to zero yields the solution. This indicates
\begin{equation}\label{label1}
 \left(      \sum_{k=1}^{N_{\text{MC}}}{\Psi \left(\sS_t^k,a_t^k\right)\Psi^\top \left(\sS_t^k,a_t^k\right)}    \right) \vec{\mathbf{W}}_t^*  
\end{equation}
would equate to
\begin{equation}\label{label2}
	\sum_{k=1}^{N_{\text{MC}}}{\Psi\left(\sS_t^k,a_t^k\right)\left(\mathcal{R}_t\left(\sS_t^k,a_t^k,\sS_{t+1}^k\right) +\gamma\max_{a_{t+1}\in\mathcal{A}}Q_{t+1}^\star\left(\sS_{t+1}^k,a_{t+1}\right)\right)},
\end{equation}
for   \eqref{set2} are obtained via \eqref{label1} and \eqref{label2}. 
Implementation is intuitive as well, by utilizing $S_{nm}^{ (t) }$, $M_n^{ (t) }$  and vector $\vec\Psi\left(\sS_t,a_t\right)$ to compute $\vec{\bf W}_t$ and learn the Q-function implied by the input data (the data that is fed to the algorithm) backward recursively with a terminal condition. 
Then calculate the matrix 
\begin{equation}
	\mathbf U_{W} \left(t,\sS_t \right) = 
	\begin{bmatrix} \mathbf U_W^{0,k}\left(t,\sS_t \right) \\  
		\mathbf U_W^{1,k}\left(t,\sS_t \right) \\ \mathbf U_W^{2,k} \left(t,\sS_t \right)
	\end{bmatrix}
	= \bf{W}_t \Phi_t \left(t,\sS_t \right),
\end{equation}
and read off the rows of this matrix to as $\mathbf U_W^{(1)}$, $\mathbf U_W^{(2)}$ and $\mathbf U_W^{(3)}$, 
along with a terminal condition specified for  $Q_T^\star\left(\sS_T,a_T^\star=0\right)=-\Pi_T\left(\sS_T\right)-\lambda \mathbb{V}\text{ar}\left[\Pi_T\left(\sS_T\right)\right]$ and $a_t^*(\sS_t)= - \frac{U_W^{(1)}(t,\sS_t)}{U_W^{(2)}(t,\sS_t)}$.
Therefore, one can obtain the optimal Q-function via
\begin{equation}
	Q_t^\star (\sS_t,a_t^\star )=\mathbf U_W^{(0)}(t,\sS_t)+ a_t^\star \mathbf U_W^{(2)}(t,\sS_t) +\frac{1}{2}(a_t^\star)^2\mathbf U_W^{(2)}(t,\sS_t).
\end{equation}
For further details we refer to~\cite{halperin2019qlbs,halperin2020qlbs,dixon2020applications}, as it could be observed, the approach is relatively straight-forward. 
However, let us reemphasize again the importance of the availability of the data that is suitable for the problem at hand.

\section{Numerical Illustrations}

Our methodology fosters the need and benefits of data-driven approaches. 
Fitted Q-Iteration assumes no knowledge of the model, it only depends on the sample data. 
In our setting, we first are in need of a predetermined hedging strategy since we aim to foresee the quoted exchange rate process. 
However, in doing so, we provide neither optimal nor sub-optimal action values for the algorithm to learn and produce optimality. 
This amounts to the fact that we aim to measure if the extended QLBS model in facing such quoted exchange rate processes is capable of delivering strategies that have less transaction costs and option prices that converge well to the bench-marking price (the fair price the large trader considers). 
Through this, we aim to make sure the large trader collects the premium that motivates him in issuing such contingent claims yet providing the large trader with a cost efficient hedging structure that also produce a convergent option price to ask. Though we set $\lambda \in \R^+$ in general case, in our study we set $\lambda = 0.001$ similar to~\cite{halperin2019qlbs}  

In this part, we sample the predetermined hedging strategy, however, we are aware of the fact that in reality there could be budget constraints or similar considerations.
In our study, we relax such constraints and consider the initial parameters provided in \tabref{input_impact}. 
We actually not only price an option, but  empirically analyze the effects of  parameters that we include through studies~\cite{roch2011liquidity,saito2017hedging}, namely; thinness and market impact, along with the effects of the size of transaction. 
Once generated optimal action policies, the large trader should be able to calculate the risks and his hedging costs given $N_{\text{MC}}$ Monte Carlo paths.  

\begin{table}[htbp]
	\centering
	\caption{Input Parameters}
	\begin{tabular}{cccccccccc}
		$F_0$ & $K$   & $\beta_t$ & $\mu$ & $\sigma$ & $r^\text{d}$ & $r^\text{f}$ & $\tau$   & $T$   & $N_{\text{MC}}$ \\
		\midrule
		2.4   & 3     & [0,1) & 0.05  & 0.05  & 0.05  & 0.00  & 0.3   & 30    & 1000 \\
	\end{tabular}%
	\label{input_impact}%
\end{table}%

The setting is briefly as follows. The large trader observes the unaffected exchange rate process up until the initiation of the issue, and devises a predetermined hedging strategy $\{u_t \}_{t=0}^T$ given  $\{F_t \}_{t=0}^T$. 
Recall  \eqref{data_av}, in which we observe state variables. We then provide the state variables computed for the quoted exchange rate process, $\{\FF\}_{t=0}^T$. 
Yet, we should draw attention to the fact that the trader calculates a predetermined hedging strategy given $N_{\text{MC}}$, and the algorithm delivers   $\{a_t^* \}_{t=0}^T$. 
Besides, recall that the quoted exchange rate process affected by the optimal action values $\{a_t^* \}_{t=0}^T $ is the implied exchange rate process, $\{\tilde{F}_t\}_{t=0}^T$.  
All option prices are calculated based on the following relation:
\begin{equation}
		C_0^{\text{QLBS}}  = -e^{-r^d T}Q_T(F_{T^+})
\end{equation}
For our illustrations, we assume intuitive predetermined hedge strategies given the initial unaffected exchange rate, the characteristics of the put option and our reformulation that could become obsolete with much larger values since $F_0$ is taken 2.4 in \tabref{input_impact}. 
We also note that, again, we prefer small-sample inference, therefore producing 50 runs of the extended QLBS under market impacts model prices. 
We, similarly to the QLBS model, prefer to work under physical domestic measure $\mathbb{P}^\text{d}$, 
yet only difference is that we simulate the unaffected exchange rate process through the foreign exchange equivalent of BSM, Garman-Kohlhagen model~\cite{garman1983foreign}. 
Let $\mu$ be the drift and $\sigma$ be the volatility of the unaffected exchange rate process,
\begin{equation}
dF_t = \mu F_t dt + \sigma F_t dW_t, \quad t = 0, \ldots,T,
\end{equation} 
in which $W_t$ is a (standard) Brownian motion. 
Rather using the change of variables utilized in the QLBS model, by \eqref{log_ret} we form the state space. 

Unless stated otherwise, we utilize the arbitrarily selected inputs given in \tabref{input_impact}, where $T$ refers to the number of time step of the formulation, we let $\tau$ denote the time to maturity. 
In our setting, the time to maturity is equally divided by the number of time steps, as $T$ refers to the horizon of the problem of the learning. 
To be able to initiate $\sS_0$ we also assume that the unaffected exchange rate right before $F_0$ is set to 2.3\footnote{%
	Our calculations show that as long as  the difference between $F_{-1}$ and $F_0$ insignificant, we get viable initiations. However, changes greater that 3 to 4 percent causes instability on general.}. 
A characteristic of European option contract, as usual, is denoted by $K$ with that we refer to the strike. 
Besides, recall that the learning period is fixed at $T$ number of steps which is equal to the time to maturity we specify assuming at $t=0$ an option is issued.   

\subsection{Effects of $M_t$}

We readjust our thinness of order book parameter, $M_t$, based on similar steps given by~\cite{saito2017hedging}, in which $M_t$ ranges from 0 to 5 and initial stock price is 100. 
However, in our study we initiate our simulations for a much lower value of initial unaffected exchange rate, i.e. $F_0=2.4$. 
Therefore, initial parameters are readjusted in this case. 
If not carefully assigned such parameters, there exists a higher likelihood of generating negative quoted exchange rate process from unaffected exchange rate process. 
This is not rather from the nature of the formulation of the model but it is reinforced by the selection of input parameters.

In the first strategy devised, we assume  $\{u_t\}_{t=0}^T $ uniformly distributed over $[-1,1)$. We readjust and sample our $M_t$ parameters uniformly within bounds on $[0.01,0.03)$, $[0.03,0.07)$, $[0.07,0.10)$ and  $[0.01,0.10)$, 
and report Mean Squared Error (MSE) of the option price to ask and Average Transaction Costs of allocation for both strategies, 
namely  $L^p$ and $L^*$. 
While  $L^p$ refers to the transaction cost associated to the strategy the trader postulates,  $L^*$ refers to the transaction cost associated to the optimal strategy that generates the implied exchange rate process along with the optimal action values for the trader. 

First, let us visually compare $\{F_t\}_{t=0}^T$, $\{\FF\}_{t=0}^T$ and $\{\tilde{F}_t\}_{t=0}^T$ for $M_t \in[0.01,0.03)$ and $M_t \in [0.03,0.07)$ in \figref{[0.01,0.03)}. 
The reason is that these bounds cover vastly the above readjusted parameters. In comparison, we keep the unaffected exchange rate process the same for all $N_{\text{MC}}$ experiments pair-wise and randomly select the plots to observe.

\begin{figure}  
	\begin{subfigure}[c]{0.5\textwidth}
		\centering
		\includegraphics[width=0.8\linewidth]{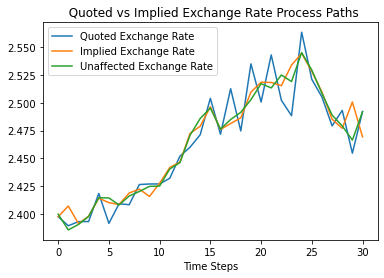}
		\caption{$M_t \in[0.01,0.03)$}
		\label{fig:comp1}
	\end{subfigure}
	\begin{subfigure}[c]{0.5\textwidth}
		\centering
		\includegraphics[width=0.8\linewidth]{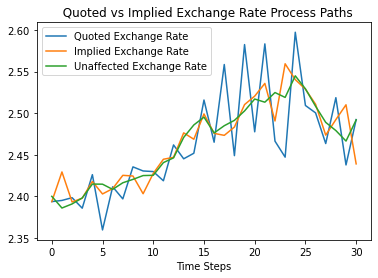}
		\caption{$M_t \in [0.03,0.07)$}
		\label{fig:comp1*}
	\end{subfigure}
	\begin{subfigure}[c]{0.5\textwidth}
		\centering
		\includegraphics[width=0.8\linewidth]{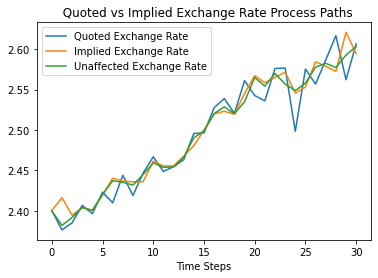}
		\caption{$M_t \in[0.01,0.03)$}
		\label{fig:comp2}
	\end{subfigure}
	\begin{subfigure}[c]{0.5\textwidth}
		\centering
		\includegraphics[width=0.8\linewidth]{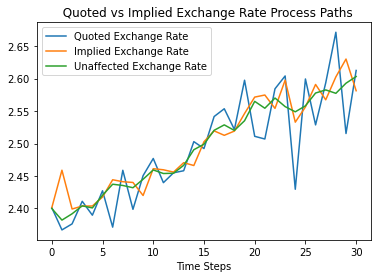}
		\caption{$M_t \in [0.03,0.07)$}
		\label{fig:comp2*}
	\end{subfigure}
	\caption{Comparison of $M_t \in[0.01,0.03)$ and $M_t \in[0.03,0.07)$ }
\label{[0.01,0.03)}
\end{figure}

In  \figref{[0.01,0.03)}, we first observe that higher values of thinness of the order books results in larger swings in quoted exchange rate processes, we could observe noticeable differences between $\{\FF\}_{t=0}^T$ and $\{\tilde{ F}_{t^+}\}_{t=0}^T$. 
Although our purpose is not to measure the degree of steepness and the range that it spans, we should note that such large movements in the quoted exchange rate process could lead to higher transaction costs compared to the transaction costs that the large trader incurs for the case of implied exchange rate process. 
Recall for this pair-wise comparison we eliminate the effect of different unaffected exchange rate processes.
 
Another point is that given $\{\tilde{ F}_{t^+}\}_{t=0}^T$ we observe smoother movements in all cases in comparison to $\{\FF\}_{t=0}^T$. 
Therefore our reformulation could lead to smoother market movements, thus the large trader avoiding regulatory authorities. 
The reason for this amounts to the possibility of market misconduct as the large trader is defined as a participant that is able to steer the foreign exchange markets. 
For a perspective on regulations and the large trader, we refer to~\cite{gastineau1991large}. 
However, we should mention that our reformulation is not on market manipulation and the large trader in our proposed model carries no hidden motives as such. 
Our approach, visually, so far proves that smoother movements are possible once we plug in the optimal action values into our exchange rate dynamics rather than following the postulated or predetermined hedging strategy. 
In order not to cause repetitions, we  present such a visual comparison for the case of $\{u_t\}_{t=0}^T \in [-1,1)$ and for the bounds $M_t \in[0.01,0.03)$  and $M_t \in [0.03,0.07)$.  

In our setting, the large trader not only seeks alternative hedging strategy, but also aims to reduce the transaction costs. 
Our purpose is to calculate the option price the large trader considers fair as well as devising a cost-efficient strategy, i.e. optimal action values. 
Recall that the large trader first calculates the price to ask based on the predetermined hedging strategy. 
The large trader then implements  Monte Carlo simulation, utilizing the quoted exchange rate process generated by the postulated hedge strategy $\{u_t\}_{t=0}^T$. 
Based on~\cite{grau2008applications}, we utilize a proportional transaction cost factor $\kappa=0.01$ defined for $ t=0,\ldots,T$ as
\begin{equation}\label{transaction_cost}
	L_t =   \kappa \abs{\FF \Delta a_t  },
\end{equation}
which leads to average $L^p$, the average of $L_t$ for $t=0,\ldots,T$, and similarly for the implied exchange rate $L^{*}$ as well.

\begin{table}%[htbp]
	\centering
	\caption{Results for   $\{u_t\}_{t=0}^T \subseteq [-1,1)$ }
	\begin{tabular}{cccc}
		$M_t$ & MSE   &  Average  $L^p$ & Average $L^*$ \\
		\midrule
		$[0.01, 0.03)$ & 0.000000001980 & 0.480035256843 & 0.182249030442 \\
		$[0.03, 0.07)$ & 0.000000000030 & 0.480376371018 & 0.154875148169 \\
		$[0.07, 0.10)$ & 0.000000000058 & 0.480467501182 & 0.168214789041 \\
		$[0.01, 0.10)$ & 0.000000000037 & 0.479848646807 & 0.234253834963 \\
	\end{tabular}%
	\label{impact_tab1}%
\end{table}%

In \tabref{impact_tab1}, comparative metrics are presented, along with average of total cost of transactions incurred during the horizon of the problem, i.e. $T=30$. 
In order, the results are  reported $M_t \in [0.01,0.03) $, $M_t \in [0.03,0.07) $, $M_t \in [0.07,0.10) $ and $M_t \in [0.01,0.10) $. 
Note that the last row corresponds to the case of  $M_t \in [0.01,0.10) $in order to investigate how well our reformulation converges to the fair price of the large trader while in larger spread; 
%We obtain  convergences to the fair price the large trader considers. 
besides, for the given parameters, average cost of transaction is noticeably lower for the optimal action values.

\begin{figure} 
	\begin{subfigure}[c]{0.5\textwidth}
		\centering
		\includegraphics[width=0.8\linewidth]{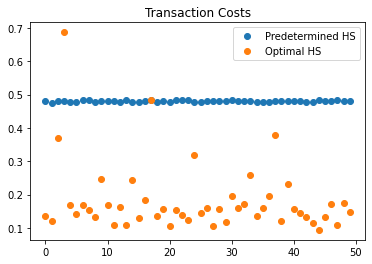}
		\caption{for $M_t \in [0.01,0.03)$}
		\label{fig:cost1}
	\end{subfigure}
	\begin{subfigure}[c]{0.5\textwidth}
		\centering
		\includegraphics[width=0.8\linewidth]{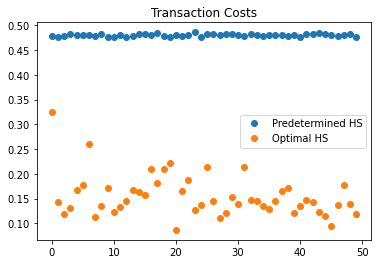}
		\caption{$M_t \in [0.03,0.07) $}
		\label{fig:cost2}
	\end{subfigure}
	\begin{subfigure}[c]{0.5\textwidth}
		\centering
		\includegraphics[width=0.8\linewidth]{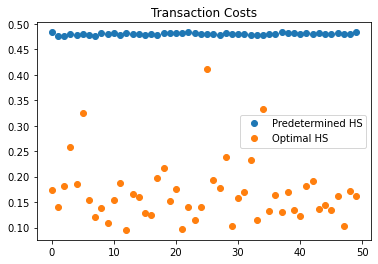}
		\caption{$M_t \in [0.07,0.10) $}
		\label{fig:cost3}
	\end{subfigure}
	\begin{subfigure}[c]{0.5\textwidth}
		\centering
		\includegraphics[width=0.8\linewidth]{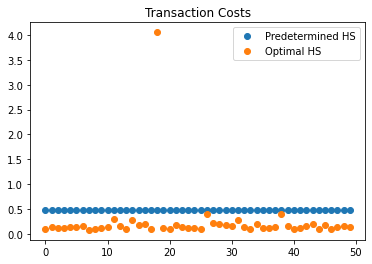}
		\caption{$M_t \in [0.01,0.10) $}
		\label{fig:cost4}
	\end{subfigure}
	\caption{for   $\{u_t\}_{t=0}^T \subseteq [-1,1)$}
	\label{for_min_comparison}
\end{figure}

Yet, merely reporting such values could be misleading as unlikely outliers could disturb the average values with a much sharper skew from the expected average. 
%For this we plot \figref{for_min_comparison}. 
We observe in \figref{for_min_comparison} that transaction costs virtually almost stay under the cost of predetermined hedging strategy, and we  remark  the effects of randomness our reformulation incurs. 
We, nonetheless aiming for much lower costs, should suggest caution as there could exist several outliers. 

\begin{figure} 
	\begin{subfigure}[c]{0.5\textwidth}
		\centering
		\includegraphics[width=0.7\linewidth]{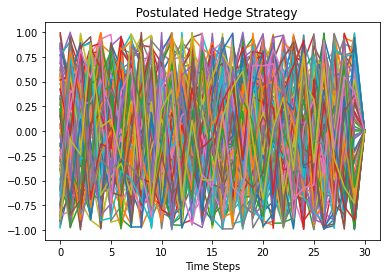}
		\caption{ $M_t \in [0.01,0.03)$}
		\label{fig0.01,0.03:4}
	\end{subfigure}
	\begin{subfigure}[c]{0.5\textwidth}
		\centering
		\includegraphics[width=0.7\linewidth]{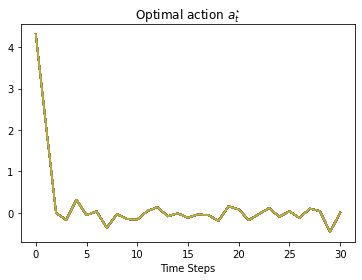}
		\caption{ $M_t \in [0.01,0.03)$}
		\label{fig0.01,0.03:5}
	\end{subfigure}
	\begin{subfigure}[c]{0.5\textwidth}
		\centering
		\includegraphics[width=0.7\linewidth]{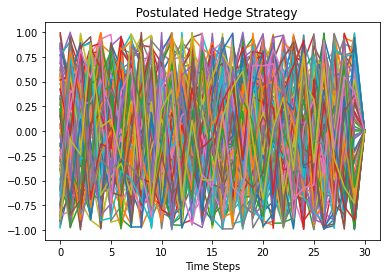}
		\caption{$M_t \in [0.03,0.07) $}
		\label{fig0.03,0.07:4}
	\end{subfigure}
	\begin{subfigure}[c]{0.5\textwidth}
		\centering
		\includegraphics[width=0.7\linewidth]{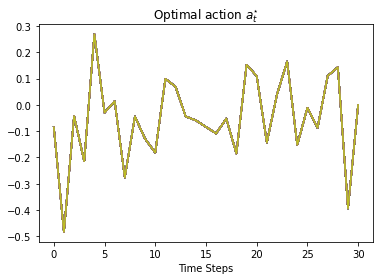}
		\caption{$M_t \in [0.03,0.07) $}
		\label{fig0.03,0.07:5}
	\end{subfigure}
	\caption{Comparison of Postulated and Optimal Action  Values}
	\label{sum_imp_1}
\end{figure}
%----end of it 

In \figref{sum_imp_1}, we compare the postulated strategy that leads to the process we define as the quoted exchange rate process with the optimal hedging decisions that we come up with. 
We see that in \figref{fig0.01,0.03:4} and \figref{fig0.03,0.07:4} the ranges are within the bounds of the predetermined hedging strategy. 
However, \figref{fig0.01,0.03:5} and \figref{fig0.03,0.07:5} present us with the fact that optimal action values know no such boundaries. 
Therefore we show that our approach is able to calculate a more efficient strategy while being under no constraints of formerly postulated hedging strategies.

We now carry out similar experiments on  $M_t \in [0.01,0.03)$, $M_t \in [0.03,0.07)$, $M_t \in [0.07,0.10)$ and $M_t \in [0.01,0.10)$, now with $\{u_t\}_{t=0}^T \subseteq [-1.5,1.5)$. Results are reported in \tabref{tab:min1.5}.
We observe that in each case, $L^p >  L^*$ as expected. 
Higher values of $M_t$ return a higher mean squared errors, while providing us with less $L^*$. 
However, even in a wider range $\{u_t\}_{t=0}^T \subseteq [-1.5,1.5)$, we conclude that our reformulation converges quite well to the fair price the large trader asks, along with a cost-efficient hedging decisions.

\begin{table}%[htbp]
	\centering
	\caption{Results for   $\{u_t\}_{t=0}^T \subseteq [-1.5,1.5)$}
	\begin{tabular}{cccc}
		$M_t$ & MSE   &  Average  $L^p$ & Average $L^*$ \\
		\midrule
		$[0.01, 0.03)$ & 0.000000000049 & 0.719687524861 & 0.227291937967 \\
		$[0.03, 0.07)$ & 0.000000000049 & 0.719565886686 & 0.263730890664 \\
		$[0.07, 0.10)$ & 0.000000000135 & 0.719958391758 & 0.257430825626 \\
		$[0.01, 0.10)$ & 0.000000000065 & 0.719581585817 & 0.212550514763 \\
	\end{tabular}%
	\label{tab:min1.5}%
\end{table}%

\begin{figure}%[H]
	\begin{subfigure}[c]{0.5\textwidth}
		\centering
		\includegraphics[width=0.8\linewidth]{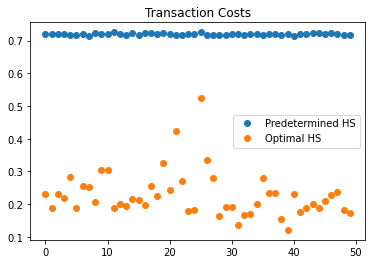}
		\caption{for $M_t \in [0.01,0.03)$}
		\label{fig:cost5}
	\end{subfigure}
	\begin{subfigure}[c]{0.5\textwidth}
		\centering
		\includegraphics[width=0.8\linewidth]{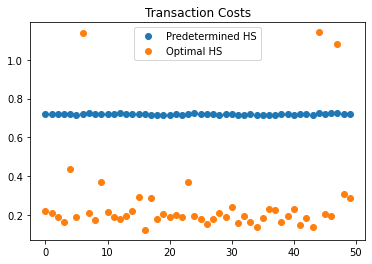}
		\caption{$M_t \in [0.03,0.07) $}
		\label{fig:cost6}
	\end{subfigure}
	\begin{subfigure}[c]{0.5\textwidth}
		\centering
		\includegraphics[width=0.8\linewidth]{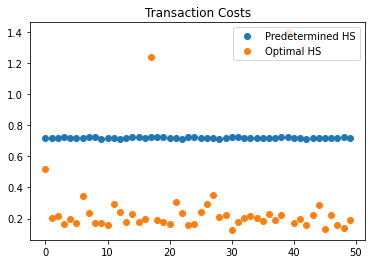}
		\caption{$M_t \in [0.07,0.10) $}
		\label{fig:cost7}
	\end{subfigure}
	\begin{subfigure}[c]{0.5\textwidth}
		\centering
		\includegraphics[width=0.8\linewidth]{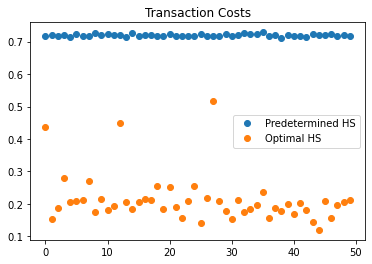}
		\caption{$M_t \in [0.01,0.10) $}
		\label{fig:cost8}
	\end{subfigure}
	\caption{for   $\{u_t\}_{t=0}^T \subseteq [-1.5,1.5)$}
	\label{key11}
\end{figure}

In \figref{key11}, we observe the corresponding transaction cost for each case as well. 
Recall that the structure is sequential, therefore we report the summation of expenditure of sequential transaction. 
However, through our analysis, there is a likelihood of outliers in much larger values of $M_t$. 
Several points should be made in explaining such outliers. First note that by \eqref{transaction_cost} and opposite signs of repositioning, i.e. buy/sell decisions, consecutively could lead to a much greater proportional transaction cost for the time step in-question. 
Besides, in several cases, optimal action values could take relatively higher values, and combining such events might lead to  higher expenditure on transactions. 
Another factor is the selection of the postulated strategy, i.e. the sampling.  

As for the effect on outliers stemming from the postulated hedge strategy, we conduct  similar experiments with counter intuitive ranges. 
Such ranges, intuitively, are unlikely to be chosen given our default parameters. 
Though optimal action values consider no limits on the values they could take, enforcing the algorithm via an unlikely selection of ranges could increase the possibility of outliers and its frequency. 

We provide \tabref{tab_unlikely} and present results for $\{u_t\}_{t=0}^T \subseteq [-2,2)$, $\{u_t\}_{t=0}^T \subseteq [-2,0)$, $\{u_t\}_{t=0}^T \subseteq [0,2)$
and $\{u_t\}_{t=0}^T \subseteq [-5, 5)$ in order. 
\tabref{tab_unlikely} shows that even in case of unlikely parameters, extended QLBS model converges to the expected fair price with which the large trader aims to collect premiums. 
Only in  $\{u_t\}_{t=0}^T \subseteq [-2,0)$
and $\{u_t\}_{t=0}^T \subseteq [0,2)$, 
higher costs associated appear. 
We conclude the reason for this is  letting the postulated strategy either on not take negative or positive values. 
Again we remark that such selections are unlikely since the forward perspectives of the underlying might require to take buy and sell simultaneously rather than allowing to buy up or sell down.

\begin{table}%[htbp]
	\centering
	\caption{Comparison of Counter-intuitive Postulated Hedge Strategies}
	\begin{tabular}{cccc}
		$u_t$ & MSE   &  Average  $L^p$ & Average $L^*$ \\
		\midrule
		$[-2,2)$ & 0.000000000117 & 0.958684889528 & 0.431902058479 \\
		$[-2,0)$ & 0.000000000061 & 0.468986820672 & 0.590129089448 \\
		$[0,2)$ & 0.000000000063 & 0.516246305467 & 0.652188718604 \\
		$[-5,5)$ & 0.000000001874 & 2.402780944608 & 1.068419632085 \\
	\end{tabular}%
	\label{tab_unlikely}%
\end{table}%

\begin{figure}
	\begin{subfigure}[c]{0.5\textwidth}
		\centering
		\includegraphics[width=0.8\linewidth]{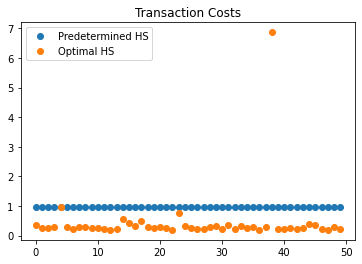}
		\caption{$\{u_t\}_{t=0}^T \subseteq [-2,2)$}
		\label{unl1}
	\end{subfigure}
	\begin{subfigure}[c]{0.5\textwidth}
		\centering
		\includegraphics[width=0.8\linewidth]{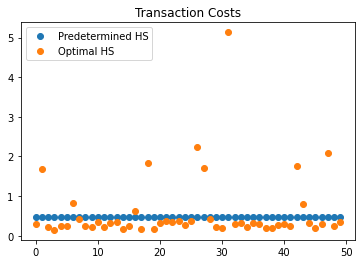}
		\caption{$\{u_t\}_{t=0}^T \subseteq [-2,0)$}
		\label{unl2}
	\end{subfigure}
	\begin{subfigure}[c]{0.5\textwidth}
		\centering
		\includegraphics[width=0.8\linewidth]{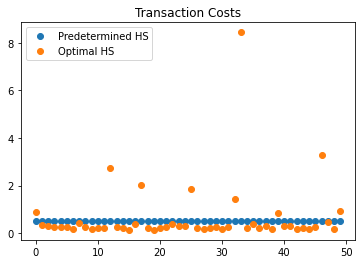}
		\caption{$\{u_t\}_{t=0}^T \subseteq [0,2)$}
		\label{unl3}
	\end{subfigure}
	\begin{subfigure}[c]{0.5\textwidth}
		\centering
		\includegraphics[width=0.8\linewidth]{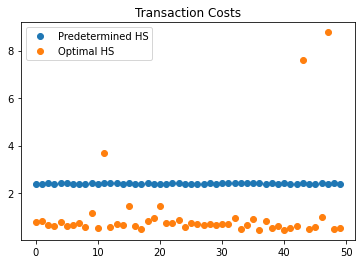}
		\caption{$\{u_t\}_{t=0}^T \subseteq [ -5, 5)$}
		\label{unl4}
	\end{subfigure}
	\caption{Unlikely Parameters for Postulated Hedge Strategy}
	\label{unlikely_plots}
\end{figure}

However, commenting on such values alone in this case is highly misleading. Therefore, in \figref{unlikely_plots} we observe that, in general, $L^*$ has lower values dispersed. 
However, noting the outliers' magnitude that the average for our sample increases easily. 
Therefore, our reformulation requires sensitive consideration for the initial parameters involved in the postulated hedge strategy. 

\subsection{Effects of $\beta_t$}

Another parameter that in embedded in the quoted exchange rate dynamics is the market impact parameter, we uniformly sample from the range $[0,1)$ following the literature in~\cite{roch2011liquidity}. 
We measure the effects of higher and lower market impact while we keep $M_t \in [0.01,0.03)$, and compare the option price generated by our proposed model to the fair price the large trader expects. 
We follow through such experiments with the strategy $\{u_t\}_{t=0}^T$ postulated by the large trader that uniformly distributes on $[-1.5,1.5)$. 
The reason for our selection of $M_t$ for this range is to ensure that thinness of order book parameter does not interfere with the in-question effects of the parameter in-question. 

\begin{table}%[htbp]
	\centering
	\caption{Effects of $\beta_t$% %\in[0,0.4)$ and $\beta_t \in [0.7,1)$ 
	}
	\begin{tabular}{cccc}
		$\beta_t$ & MSE   &  Average  $L^p$ & Average $L^*$ \\
		\midrule
		$[0,0.4)$ & 0.000000000021 & 0.479730599078 & 0.287987013094 \\
		$[0.7,1)$ & 0.000000000063 & 0.479651389605 & 0.240905881121 \\
	\end{tabular}%
	\label{d11}%
\end{table}%

\begin{figure}%[H]
	\begin{subfigure}[c]{0.5\textwidth}
		\centering
		\includegraphics[width=0.8\linewidth]{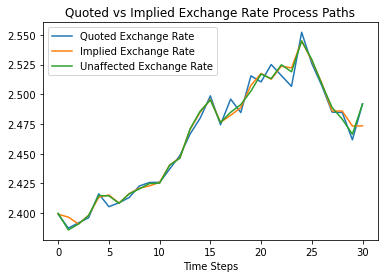}
		\caption{$\beta_t\in[0,0.4)$}
		\label{fig:61}
	\end{subfigure}
	\begin{subfigure}[c]{0.5\textwidth}
		\centering
		\includegraphics[width=0.8\linewidth]{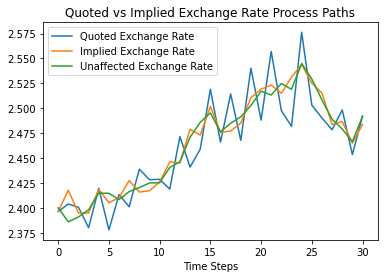}
		\caption{ $\beta_t \in [0.7,1)$}
		\label{fig:62}
	\end{subfigure}
	\begin{subfigure}[c]{0.5\textwidth}
		\centering
		\includegraphics[width=0.8\linewidth]{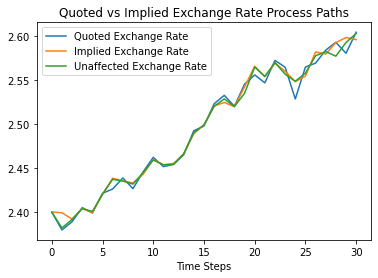}
		\caption{$\beta_t \in [0,0.4)$}
		\label{fig:1661}
	\end{subfigure}
	\begin{subfigure}[c]{0.5\textwidth}
		\centering
		\includegraphics[width=0.8\linewidth]{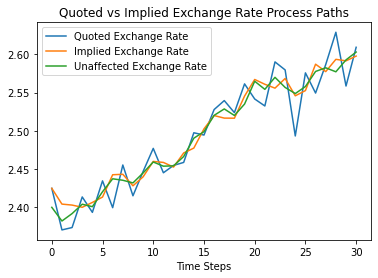}
		\caption{ $\beta_t \in [0.7,1)$}
		\label{fig:1662}
	\end{subfigure}
	\caption{Comparison of Effects of $\beta_t$}
	\label{beta_comp2}
\end{figure}

Results of numerical experiments for lower and higher levels of market impact parameter can be found in \tabref{d11}. 
As expected, and confirmed by \figref{beta_comp2}, MSE values increase with respect to $\beta_t$, andverify once again that our reformulation generates a hedging strategy in terms of cost efficiency.

In \figref{beta_comp2}, we compare the results obtained by $\beta_t \in [0,0.4)$ and $\beta_t \in [0.7,1)$. 
Pairs in panels such as \figref{fig:61} and \figref{fig:62} or \figref{fig:1661} and \figref{fig:1662} utilize the same unaffected paths of the exchange rate process. 
Therefore, basically, on the left panel we present results for $\beta_t \in [0,0.4) $ and on the right for $\beta_t \in [0.7,1)$. 
Consider \figref{fig:62}, we clearly observe the effect of the market impact parameter, leading much steeper behavior compared to the quoted exchange rate process presented in \figref{fig:61}. 
Similar arguments apply to others as well. 
In each of the plots, the quoted exchange rate exhibit divergence, while for the lower values of $\beta_t$, less divergence seems the case. 
Therefore, in preview of hedging decisions, the importance of market impact parameter persists.

\begin{figure}%[H]% 
\begin{subfigure}[c]{0.5\textwidth}
		\centering
		\includegraphics[width=0.9\linewidth]{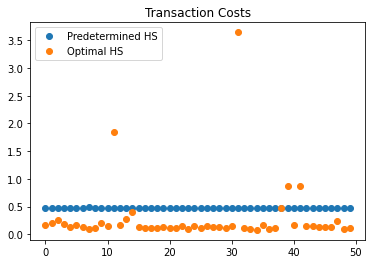}
		\caption{For $\beta_t \in [0,0.4)$}
		\label{(0,0.4)}
\end{subfigure}
\begin{subfigure}[c]{0.5\textwidth}
	\centering
	\includegraphics[width=0.9\linewidth]{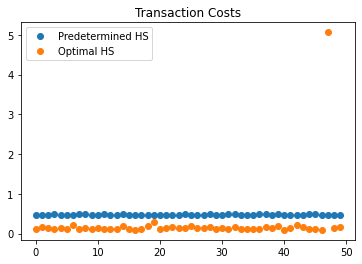}
	\caption{For $\beta_t \in [0.7,1)  $ }
	\label{(0.7,1)}
\end{subfigure}
\caption{Transaction Costs for $\beta_t \in [0,0.4)$ and $\beta_t \in [0.7,1)$ }
\label{beta_comp}
\end{figure}

Lastly, we present the dispersion of transaction costs for both cases of $\beta_t$ in \figref{beta_comp}. 
We observe, again, outliers are highly greater in magnitude. 
Though throughout our study, we prefer to work with small-sample inference, we suggest increasing the number of runs to be better prepared in advance. 
However, given our sampling, we conclude and argue that our reformulation converges well while delivering a cost-efficient hedging strategy.

\section{Conclusion and Outlook}	
	
In this paper, through reinforcement learning, we reformulate the existence as well as the perspectives of a large agent by extending the QLBS model that takes its roots in reinforcement learning. 
Considering the literature on option pricing under market (i.e. price) impact models; 
our proposed model offers a great deal of flexibility and applicability in regard to both devising hedging strategies and being a candidate as a benchmark model.

Flexibility comes from the fact that we keep the data-driven perspectives of the QLBS model, while allowing complex effects at the same time. 
Therefore, any underlying model to simulate and test could be used. 
This contributes to the usability of our proposed model as a benchmark model in formulating and devising hedge strategies along with theoretical option prices.

One other advantage of this current study is that predetermined hedge strategy structures the quoted exchange rate process, and optimal hedge strategy leads to the implied exchange rate process. 
One can tune any sort of strategy that is tailored to the needs and could have a less costly hedging strategy.

We are also aware of a simplification we assumed, i.e. no delay given the impact: 
in the proposed model the time between $F_t$ and $\FF$ is negligible. 
This could be considered unrealistic, however, given our assumption of anticipation of the movements of the large agent, we find it acceptable. 
Nonetheless, this should be relaxed in a further study. 
In doing so, we aim to consider functions that could introduce delays in the passage of time as well as in rewards. 
Another possible further study is to include other market participants that build up along with the large agent or enter into opposite transactions, as this could possibly be an interesting set-up to observe.

In essence, the proposed model intact with embedded parameters that shape and alter the exchange rate dynamics converges well to the prior fair price concept of the agent. 
The reformulation both manages convergence given MSE values and in doing so also is capable of producing much more effective hedging values based on our finding. 
However, we, expectedly, have some simplifications in the pursue of our reformulation. 
Though we stand by the sampling $\beta_t \in [0,1)$ following the literature on~\cite{roch2011liquidity,saito2017hedging}, we suggest caution in regard to thinness of order book, $M_t$. 
Even though we proceed with attention based on the suggestion of possibility of negative prices as discussed in~\cite{roch2011liquidity}; 
we consider the requirement of estimating the parameter in-question from the real data vital.

In a further study, we mean to address this issue along with the possible introduction of stochastic interest rates. 
Overall, we argue that our revised approach could bring practicality and a newer point of view in implementing various machine learning approaches in pricing financial derivatives. 
We believe that in the coming decades more and more types of such approaches will surface and enrich the literature on finance. 
However, a limitation of real-life data could also be an issue as higher volumes might be needed. 
For that, we put forward that the possibility of using high-frequency data rather than simulation data. We leave that question open for further research.

\section{Acknowledgment}

This study is from the first author's doctoral study at Middle East Technical University which is partially funded by a scholarship granted through Council of Higher Education of Turkey in the field of Artificial Intelligence and Machine Learning under the category of 100/2000 Doctoral Scholarships. For that we would like to extend our sincere gratitude to both Council of Higher Education of Turkey and Middle East Technical University.

\bibliography{myBiblio}
\bibliographystyle{plain}	

\end{document}